\documentclass[final,3p,times]{elsarticle}

\usepackage{amssymb}
\usepackage{amsthm}
\usepackage{amsmath}
\usepackage{multirow}
\usepackage{textcomp}
\usepackage{subcaption} 
\usepackage{siunitx}
\usepackage{epstopdf}
\usepackage{booktabs}
\usepackage[table,xcdraw]{xcolor}

\journal{xxx}
\begin{document}

\begin{frontmatter}



\title{ A Lightweight Complex-Valued Deformable CNN for High-Quality Computer-Generated Holography}

\fntext[equal]{These authors contributed equally to this work.}
\author[gz]{Shuyang Xie\fnref{equal}}

\affiliation[gz]{organization={Thrust of Microelectronics of Function Hub, The Hong Kong University of Science and Technology (Guangzhou)},
            postcode={511400},
            city={Guangzhou},
            country={China}}

\author[sc]{Jie Zhou\fnref{equal}}
\affiliation[sc]{organization={College of Electronics and Information Engineering, Sichuan University},
            postcode={610065},
            city={Chengdu},
            country={China}}

\author[gz]{Bo Xu}

\author[sc]{Jun Wang\corref{cor1}}
\ead{jwang@scu.edu.cn}
\cortext[cor1]{Corresponding author}
\author[gz]{Renjing Xu\corref{cor1}}
\ead{renjingxu@hkust-gz.edu.cn}

\begin{abstract}
Holographic displays have significant potential in virtual reality and augmented reality owing to their ability to provide all the depth cues. Deep learning-based methods play an important role in computer-generated holography (CGH). During the diffraction process, each pixel exerts an influence on the reconstructed image. However, previous works face challenges in capturing sufficient information to accurately model this process, primarily due to the inadequacy of their effective receptive field (ERF). Here, we designed complex-valued deformable convolution for integration into network, enabling dynamic adjustment of the convolution kernel's shape to increase flexibility of ERF for better feature extraction. This approach allows us to utilize a single model while achieving state-of-the-art performance in both simulated and optical experiment reconstructions, surpassing existing open-source models. Specifically, our method has a peak signal-to-noise ratio that is 2.04 dB, 5.31 dB, and 9.71 dB higher than that of CCNN-CGH, HoloNet, and Holo-encoder, respectively, when the resolution is 1920$\times$1072. The number of parameters of our model is only about one-eighth of that of CCNN-CGH.
\end{abstract}




\end{frontmatter}




\section{Introduction}

Immersive augmented reality and virtual reality (AR/VR) systems \cite{nature24, np2025, chae2025light} are gradually becoming a part of everyday life with the proliferation of wearable AR/VR display devices. This field has long attracted the attention of researchers in computer graphics, optics, and vision science, who have been particularly focused on mitigating the visual fatigue caused by such displays. Computer-generated holography (CGH) \cite{tcsvt5} is widely acknowledged as a highly desirable three-dimensional (3D) technology. It has the ability to deliver complete parallax and depth information to the human eye \cite{depth1,depth2}, thereby enhancing perceptual realism and viewer comfort. These attributes are considered to have substantial potential in the AR/VR field. CGH is a method of generating holograms by simulating diffraction models in computer \cite{holo} rather than through real optical recordings and reconstruction. Spatial light modulator (SLM) is a type of device to load these holograms for reconstruction in real life. SLMs are primarily categorized into two types: amplitude-only and phase-only. Phase-only Hologram (POH) is the dominant encoding method due to its high diffraction efficiency \cite{poh}. To obtain a POH, methods are generally divided into non-iterative and iterative approaches. Non-iterative methods, such as double phase-amplitude coding (DPAC) \cite{DPAC}, process the data once to generate a POH. In contrast, iterative methods, including Gerchberg-Saxton (GS) \cite{gs}, Wirtinger Holography (WH) \cite{wh}, and stochastic gradient descent (SGD) \cite{sgd,Zhou}, can yield higher-quality reconstructed images but require extensive computation, often iterating hundreds or thousands of times.

Recently, learning-based methods \cite{dl,nature} in CGH have garnered considerable attention for their speed and high reconstruction quality. These methods can integrate the physical wave propagation model in free space, such as the angular spectrum method (ASM) \cite{asm}, into the neural network framework, making image reconstruction both efficient and accurate. Notable frameworks like HoloNet \cite{sgd} and CCNN-CGH \cite{ccnn} demonstrate the capability to generate high-quality holograms in real time. Both frameworks utilize two networks: the first network, the phase predictor, takes the target amplitude as input to predict the phase on the target plane. This predicted phase, combined with the target amplitude, forms a complex amplitude, which is then processed using forward ASM to obtain the SLM field distribution, serving as input for the second network. This second network, the hologram encoder, generates the hologram and utilizes a backward ASM to reconstruct field, then computes the loss between the reconstructed and target amplitudes, facilitating backpropagation to update network parameters. However, this framework often need a pair or more networks to generate holograms, requires more memory to store. An alternative framework, Holo-encoder \cite{holoencoder}, directly generates holograms by inputting target amplitudes into a single network without phase predicted network. While this approach simplifies and accelerates the generation process, it typically results in poorer image quality due to its reliance solely on amplitude information. Furthermore, several studies have modified these networks \cite{Yu} to improve outcomes by incorporating Fourier transforms \cite{fourier}, wavelet transforms \cite{wavelet}, and compensation networks \cite{compensation}. However, such modifications often complicate the model and demand greater computational resources and inference time.

The diffraction process is inherently global, meaning that each pixel on the hologram can affect the image on the reconstruction plane. For neural networks, this necessitates a larger effective receptive field (ERF) to achieve better global information extraction capabilities. In traditional convolutional neural networks (CNNs), utilizing larger convolutional kernels and increasing network depth are two feasible approaches to enhance the receptive field. However, these methods significantly increase the number of network parameters and substantially prolong inference time, making it challenging to develop a real-time, lightweight hologram generation network.

In this paper, we propose a straightforward yet effective framework for generating POH using a deformable convolutional neural network to increase the flexibility of ERF that achieves superior reconstruction quality and fast inference speed compared to almost existing open-source networks. Our approach employs the complex amplitude obtained after ASM of the target amplitude as the input for our CNN, which is a complex-valued CNN based on the U-Net architecture. Although we are not the first to utilize this convolutional structure, our method for hologram generation distinguishes itself from prior complex-valued approaches. 
To capitalize on the benefits of deformable convolution, we designed complex-valued deformable convolution in the form of complex amplitudes, allowing the model to more effectively capture both local details and global phase interactions, thereby enhancing performance in hologram reconstruction. Our simulation and optical experiment results indicate that our model achieves a peak signal-to-noise ratio (PSNR) that is 2.04 dB, 5.31 dB, and 9.71 dB higher than those of CCNN-CGH, HoloNet, and Holo-encoder, respectively, at a resolution of 1920$\times$1072. Additionally, our model demonstrates a comparably fast inference speed and has a parameter count approximately one-eighth that of CCNN-CGH, effectively minimizes storage and computational requirements.

\section{Related work}
Holography was first proposed by Dennis Gabor in 1948 \cite{1948A}. Research on holographic displays has been going on for decades, and we review the works of CGH in this section.

\subsection{Holographic display}
Holographic displays are able to reproduce the entire continuous light field of a given scene through SLM regulation of incident light. This capability enables them to provide all depth cues, making them highly promising for future applications in AR \cite{ar1}, VR \cite{vr1} and head-up display \cite{headup1} applications. Typically, dynamic holographic displays \cite{holo,dyna} generally employ SLMs, like phase-only liquid crystal on silicon (LCoS) devices \cite{lcos}, in conjunction with CGH algorithms.

\subsection{Computer-generated hologram}
The concept of CGH was first proposed by Lohmann et al \cite{Lohmann}. Creating an optical hologram necessitates that the object be real, allowing the object light wave and the reference light wave to coherently superimpose on the holographic plane. This requirement makes traditional holography unsuitable for virtual objects. In contrast, CGH only requires the object light wave distribution function to generate the hologram. Additionally, CGH is less susceptible to external influences and allows for easier and more precise reproduction.

Numerous CGH generation methods have emerged in recent years. In 2015, Zhao et al \cite{Zhao}. introduced a CGH algorithm based on the angular spectrum method, which effectively reduces computational load while maintaining image quality. Additionally, models such as Kirchhoff and Fresnel diffraction \cite{Goodman} are widely used for numerically propagating wave fields. In the optimization of 3D holograms, while point-cloud \cite{point} and polygon-based \cite{polygon} sampling strategies exist, the mainstream approach is to segment the object wave into multiple layers \cite{layer}. A traditional approach to optimizing 3D holography relies on wavefront superposition. All these methods aim to facilitate the rapid generation of 3D holograms. Moreover, there are iterative techniques focused on quality enhancement, such as the improved GS method proposed by Liu et al \cite{imgs}, and the multi-depth SGD method introduced by Chen et al \cite{chenchun}.

\subsection{CGH based on deep learning}

CNNs have been widely employed in the real-time generation of holograms due to their ability to efficiently handle complex computations. Peng et al. introduced a method called HoloNet \cite{sgd}, which incorporates aberrations and light source intensity into the network's learning process. This approach aims to mitigate the impact of optical equipment mismatches on experimental results, although it does not fully account for all errors. In contrast, Choi et al. \cite{n3d} proposed CNNpropCNN, which uses captured images to train the neural network to simulate physical errors, thereby addressing a broader range of mismatches during hologram generation. For 3D hologram generation, Liang \cite{nature} used RGB-D data as input and developed a network capable of photorealistic reconstruction, effectively simulating defocus effects. Yan et al. \cite{full} utilized a fully convolutional neural network to generating multi-depth 3D holograms, which can generate multi-depth 3D holograms with 4K resolution. Additionally, Choi et al. \cite{tm} employed time-multiplexing techniques to achieve impressive defocus effects with various input data types, such as focal stacks and light fields. Zhou et al. \cite{zhou2024point} proposed a deformable convolution based on the point spread function for hologram reconstruction. Liu et al. \cite{liu20234k} proposed a diffraction model-driven neural network to generate 4K holograms, while Lee et al. \cite{lee2024holosr} demonstrated real-time high-resolution hologram generation through the use of super-resolution techniques. Additionally, other approaches have employed hologram generation networks trained with hybrid domain loss to achieve high reconstruction quality \cite{zheng2023diffraction}.

Regarding real-time capabilities, Zhong et al. \cite{ccnn} utilized complex-valued convolutions to achieve fast and high-quality holograms, which utilizes a complex-valued convolutional neural network. This model significantly reduces the number of parameters while achieving the fastest generation speed. Meanwhile, Wei et al. \cite{atten} introduced self-attention mechanism into model, achieve a high perceptive index. Qin et al. \cite{gan1} employed a complex-valued generative adversarial network to generate holograms. Although the quality of these holograms surpasses that of CCNN-CGH, both the number of parameter and processing time remain substantial. 

Unlike previous methods, our approach does not rely on a phase prediction network based on complex-valued networks. Instead, we utilize the complex-valued field propagated by ASM as input. Within our network, we incorporate deformable convolution, which addresses the limitations of traditional convolutional receptive fields found in earlier networks, thereby enhancing feature extraction capabilities.

\section{Model framework}

\begin{figure}[!htbp]
    \centering
    \includegraphics[scale=1]{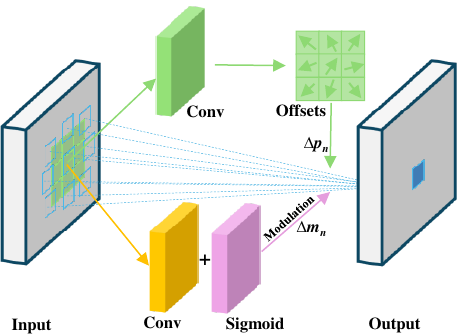}
    \caption{The illustration of 3$\times$3 deformable convolution.}
    \label{deform}
\end{figure}

\subsection{Deformable Convolution}
The traditional convolution operation involves dividing the feature map into segments that match the size of the convolution kernel, and then performing the convolution on each segment, with each segment occupying a fixed position on the feature map. However, for objects with more complex deformations, this approach may not yield optimal results. We can define the relationship between input feature $x$ and output feature $y$ with equation below\cite{deformable},
\begin{equation}
y(p_0) = \sum_{p_n \in R} w(p_n) \cdot x(p_0 + p_n),
\end{equation}
here, $R$ is a regular grid used to sample, the total of sampled values each multiplied by the weight $w$, and $p_n$ enumerates the locations in $R$.

In deformable convolution, shown in Fig. \ref{deform}, offsets are introduced into the receptive field, and these offsets are learnable. This allows the receptive field to adapt to the actual shape of objects rather than being constrained to a rigid square. Consequently, the convolutional region consistently covers the area around the object's shape, enabling effective feature extraction regardless of the object's deformation.
\begin{equation}
y(p_0) = \sum_{p_n \in R} w(p_n) \cdot x(p_0 + p_n + \Delta{p}_n),
\end{equation}
here, $\Delta{p}_n$ is the offsets at the \textit{n}-th position.
To enhance the capability of deformable convolution in controlling spatial support regions, which is introduced a modulation mechanism\cite{deformablev2}.

\begin{equation}
y(p_0) = \sum_{p_n \in R} w(p_n) \cdot x(p_0 + p_n + \Delta{p}_n) \cdot \Delta{m}_n,
\end{equation}
$\Delta{m}_n$ is the modulation scalar at the \textit{n}-th position, which is also learnable parameter. The range of modulation scalar is from 0 to 1, which is limited by sigmoid function.

\subsection{Architecture}

\begin{figure}[!t]
\centering
\includegraphics[scale=0.45]{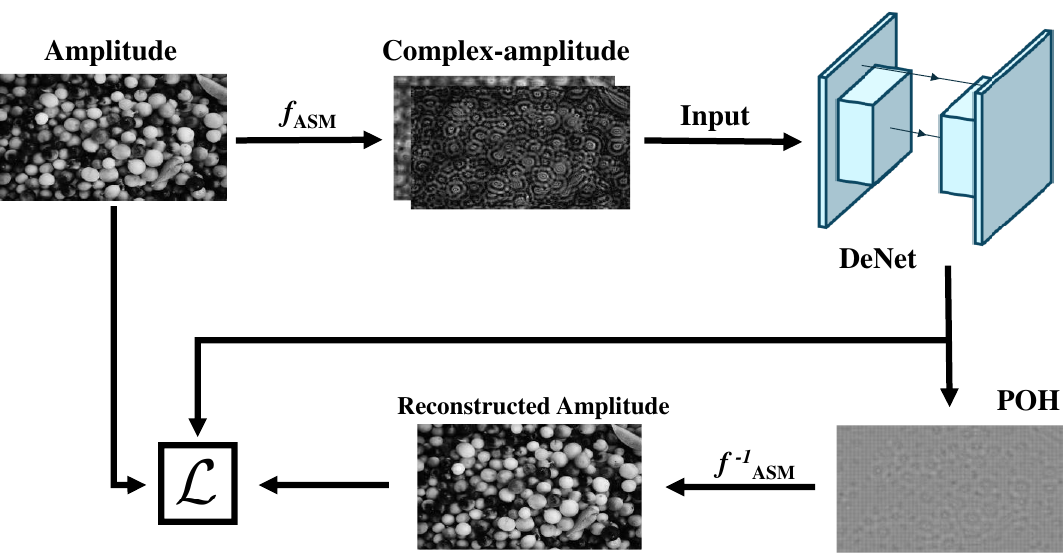}
\caption{The framework of proposed model generated POH.}
\label{framework}
\end{figure}

The framework of our model generated POH is shown in Fig. \ref{framework}. At first, the amplitude of the target image is propagated forward using the ASM to obtain the complex amplitude in the SLM plane. Complex amplitude serves as the input for our complex-valued deformable network (DeNet), allowing it to capture both amplitude and phase information. Subsequently, the POH generated by the model is propagated backward through backward ASM to reconstruct the amplitude. ASM can be expressed in equation below:

\begin{equation}
\begin{gathered}
u(\phi) = \mathcal{F}^{-1}\{\mathcal{F}\{u^{i\phi}\} H(f_x, f_y)\} \\
H(f_x, f_y) = 
\begin{cases} 
{{\rm{e}}^{i\frac{{2\pi }}{\lambda }z\sqrt {1 - {{(\lambda {f_x})}^2} - {{(\lambda {f_y})}^2}} }}, & \text{if } \sqrt {f_x^2 + f_y^2}  < \frac{1}{\lambda } \\ 
0, & \text{otherwise}
\end{cases},
\end{gathered}
\end{equation}
here, $u^{i\phi}$ is the optical field distribution, $\lambda$ is the wavelength, \textit{z} is the distance between SLM plane and target plane, $f_x$ and $f_y$ is the spatial frequencies, $\mathcal{F}$ means the Fourier transform.

The Mean Squared Error loss function ($\mathcal{L}_{MSE}$) is employed to evaluate the discrepancy between the reconstructed amplitude and its original counterpart, facilitating updates to the model parameters accordingly. Incorporating Total Variation loss ($\mathcal{L}_{TV}$) can lead to smoother phase in the hologram.

\begin{equation}
\mathcal{L} = \mathcal{L}_{MSE}(\lvert{u(\phi)}\rvert, a_{\text{target}}) + \alpha\mathcal{L}_{TV}(\phi)
\end{equation}

\begin{equation}
\mathcal{L}_{TV}(\phi) = \frac{\sum\limits_{i,j}((\phi_{i,j-1}-\phi_{i,j})^2+(\phi_{i+1,j}-\phi_{i,j})^2)}{(M-1)(N-1)}
\end{equation}here, $\lvert{u(\phi)}\rvert$ means the reconstructed amplitude and $a_{\text{target}}$ denotes the target amplitude. $\alpha$ is a weighting coefficient. During the optimization process, if $\alpha$ is too large, it can result in poor quality of the reconstructed image, while if it is too small, it may not significantly affect the smoothness of the phase. Therefore, $\alpha$ is set to $0.1\times0.1^{epoch}$ in this paper. \textit{M}, \textit{N} is the resolution of input field.

\begin{figure}[!t]
\centering
\includegraphics[scale=0.5]{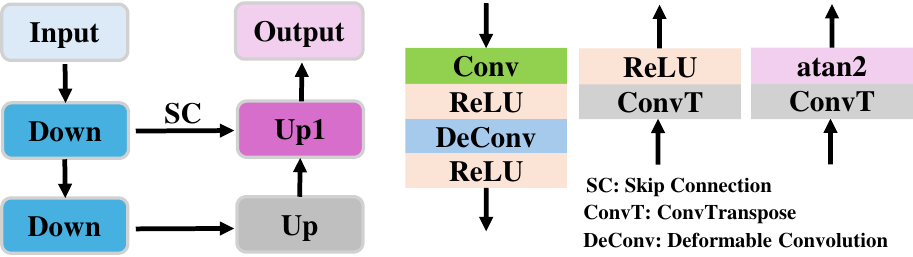}
\caption{The architecture of proposed network.}
\label{network}
\end{figure}

Fig. \ref{network} shows the detailed network architecture, which is based on U-Net. Our network utilizes complex-valued fields as inputs, rather than concatenating them together, thereby enabling feature extraction and processing to be conducted in a complex-valued format. The downsampling layer comprises a standard convolution that reduces the size of the feature map by half, along with a deformable convolution that permits the network to dynamically adjust the position of the convolution kernel during the operation. This approach significantly enhances feature representation. Both activation functions after convolution are ReLU. Upsampling layer merely employs a single deconvolutional layer to restore the feature map to the original size of the input amplitude, the activation function of the first upsampling layer is also ReLU, but that of the second layer is arctangent function, which limits the range of generated POH. And SC represents a simple addition operation.

\section{Model train and validation}

In order to validate the effectiveness of proposed model, all algorithms were implemented in Python 3.9 using the PyTorch 2.1.1 framework on a Linux workstation equipped with an AMD EPYC 7543 CPU and an NVIDIA GeForce RTX 3090 GPU.
Models were trained for 20 epochs with a batch size of 1 and a learning rate of 0.001 on the DIV2K \cite{div2k} training set, and performance was assessed on both the DIV2K and the Flickr2K validation dataset.
Holograms were generated at a resolution of 1920$\times$1072 pixels; SLM had an 8 $\mu$m pixel pitch. Optical parameters were fixed at a laser wavelength of 671 nm and a propagation distance of 200 mm.

\begin{figure}[!t]
\centering
\includegraphics[scale=0.8]{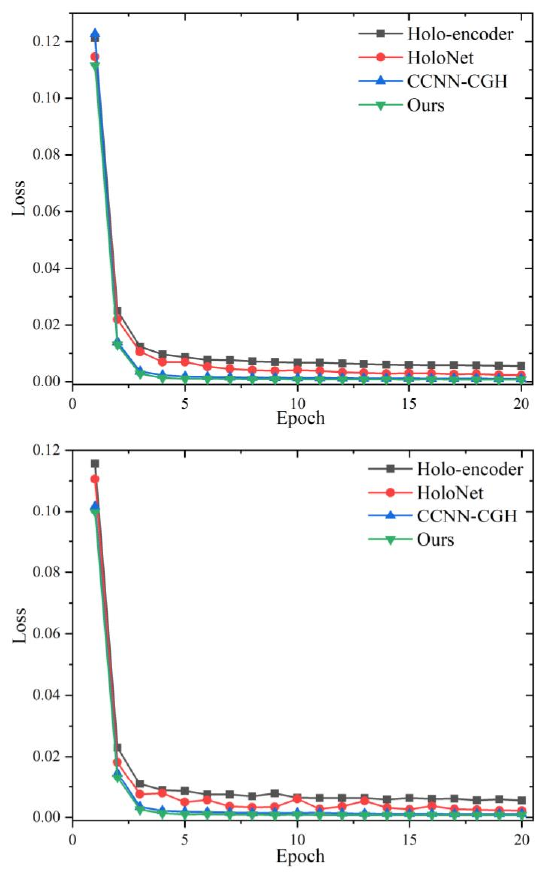}
\caption{The loss curves of train (top) and validation (bottom).}
\label{loss}
\end{figure}

Fig. \ref{loss} illustrates the loss curves for both the training and validation datasets, representing the average loss values. The results indicate that our model achieves better convergence in fewer training epochs compared to the others. Specifically, our model, along with the holo-encoder and CCNN-CGH, requires approximately 35 minutes to train for 20 epochs, whereas the HoloNet takes about 50 minutes for the same epochs. 

\begin{figure*}[h]
\centering
\includegraphics[width=6.4in]{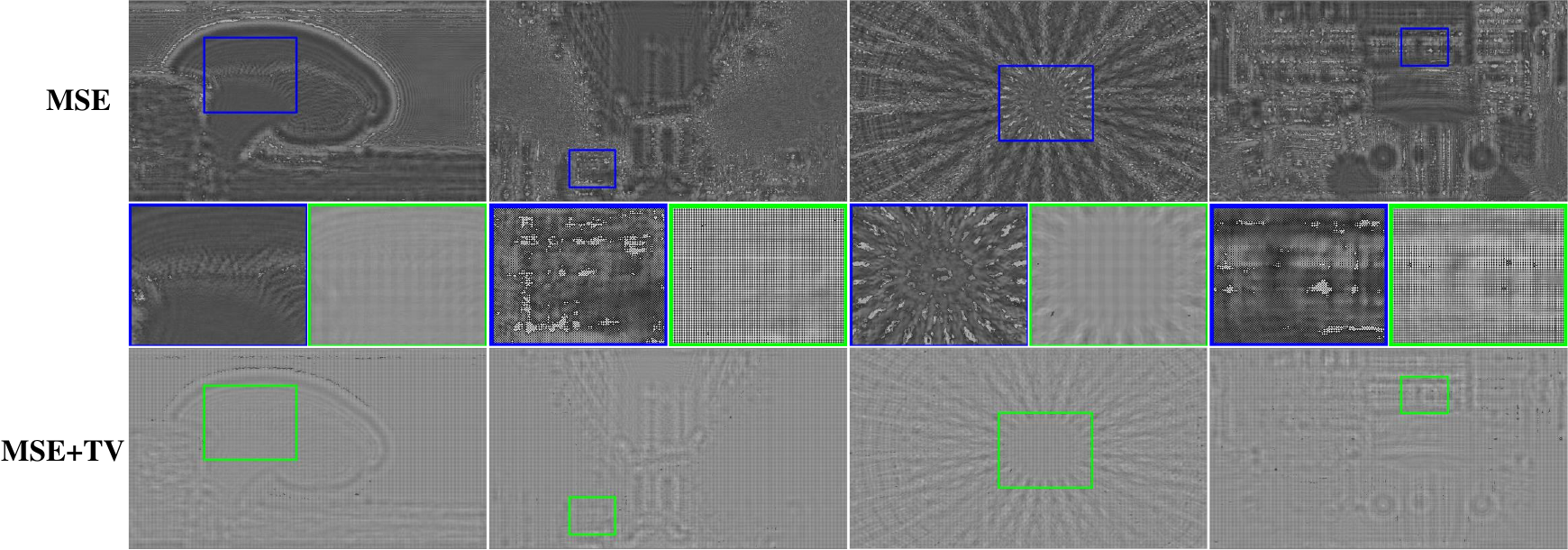}
\caption{The holograms visualization using different loss functions.}
\label{tv}
\end{figure*}

Fig. \ref{tv} illustrates the differences in holograms generated using different loss functions. Under certain initial conditions, using only MSE can result in numerous phase discontinuities. While these discontinuities may not significantly affect simulation results, they can greatly impact the quality of reconstructed images in optical experiments. By introducing TV loss, the phase continuity of the hologram is significantly improved, effectively reducing the impact of these discontinuities on optical experiments.

\begin{figure}[!t]
\centering
\includegraphics[scale=0.35]{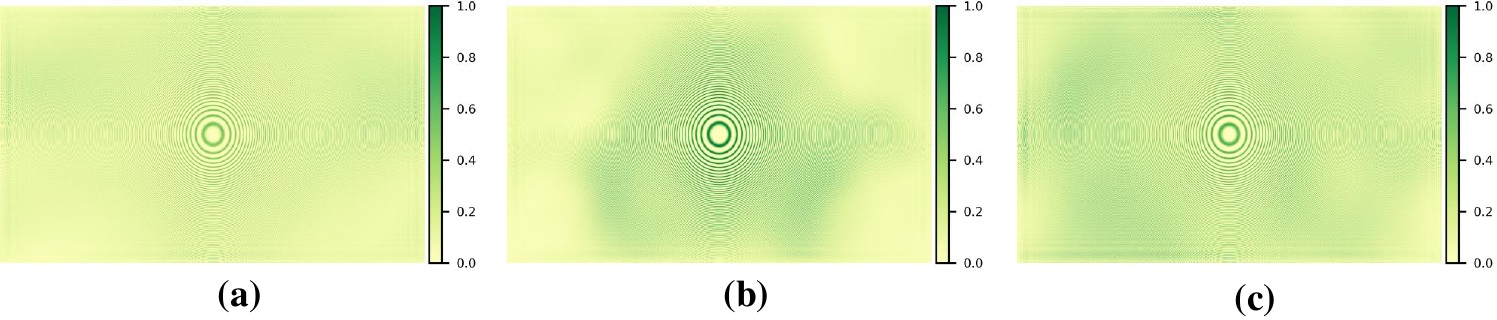}
\caption{The visualization of effective receptive field, a more extensive dark area corresponds to a larger ERF. (a) CCNN-CGH; (b) Four layers complex-valued convolution; (c) Our deformable convolution.}
\label{erf}
\end{figure}

The receptive field refers to the region in the input image that corresponds to a unit in a feature map of a CNN—that is, the area of the input that the CNN can "see". To analyze this, we compute the ERF by calculating the gradient of the central point in the output phase with respect to each point in the input image. These gradient values are then normalized for visualization. A large gradient at a particular input location indicates that it has a strong influence on the output, while a small gradient suggests minor influence. If the gradient is zero, the point lies outside the ERF and does not affect the output. As illustrated in Fig. \ref{erf}, we have visualized the ERF of three models. Fig. \ref{erf} (a) represents the CCNN-CGH, which exhibits a smaller ERF compared to our model, the influence of the peripheral regions on the central area is also relatively limited. To eliminate the impact of network depth and to validate the efficacy of deformable convolution, we replaced the deformable convolutions with a four-layer complex-valued convolution, thereby forming a five-layer complex-valued convolution as the downsampling layer. The resulting ERF is depicted in Fig. \ref{erf} (b), it remains smaller than that of our model, which is shown in Fig. \ref{erf} (c).

\begin{figure*}[!t]
\centering
\includegraphics[scale=0.55]{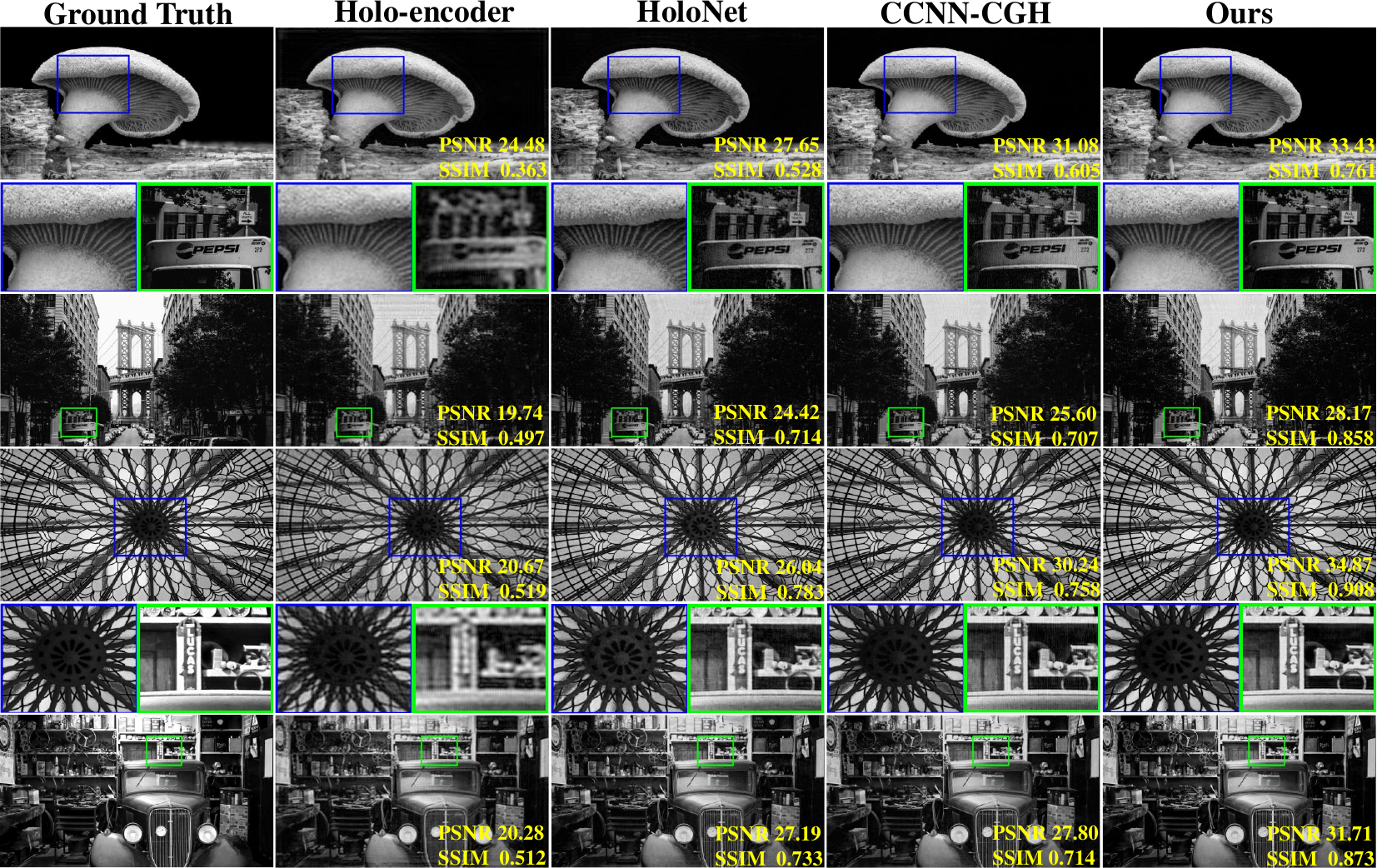}
\caption{Numerical simulation results of all evaluated methods in 1920$\times$1072 resolution.}
\label{simulation}
\end{figure*}

Since the validation set of DIV2K contains only 100 images, we conducted inference on the Flickr2K dataset to evaluate the generalization performance of our model. Table \ref{table1} present the results of numerical simulations of various models when the resolution is 1920 $\times$ 1072. To quantitatively assess reconstruction quality, we employ PSNR, Structural Similarity Index (SSIM), and floating-point operations (FLOPs) as evaluation metrics. All reported values represent the averages across the dataset and we calculate parameters and FLOPs using the thop Python package. Our model achieves superior results, with a PSNR of 33.50 dB and 33.53 dB, an SSIM of 0.921 and 0.928 on DIV2K/Flickr2K, separately. In comparison, the metrics for CCNN-CGH are 2.04 dB, 1.81 dB and 0.077, 0.065 lower than those of our model, while the performance of HoloNet and holo-encoder is comparatively weaker on both datasets. The inference speed of our model has a comparable performance compared with that of CCNN-CGH, but the reconstruction quality we achieve is significantly higher. Additionally, our model FLOPs is the lowest among all models. Fig. \ref{simulation} presents the simulated reconstructed images. The Holo-encoder performs poorly in reconstructing complex images, resulting in significant blurring. In contrast, both HoloNet and CCNN-CGH are capable of reconstructing images with greater clarity, although some noise is still present. Our model, however, achieves the best quality by reconstructing clear images with minimal noise.

Commercially available SLMs come in various pixel pitches, including 3.74 $\mu$m and 6.4 $\mu$m. To validate the effectiveness of our model, Table \ref{table2} presents the results of numerical simulations of various pixel pitches at a resolution of 1920 $\times$ 1072. Holo-encoder and HoloNet remain the two lowest-performing models. Our model still achieves the best simulation results, with a PSNR of 33.15 dB, 33.45 dB and an SSIM of 0.899, 0.910. This represents a significant improvement compared to the CCNN-CGH model, which shows a lower PSNR of 1 dB, 1.78 dB and an SSIM of 0.036, 0.062 less than ours.

\begin{table}[!t]
\centering
\caption{Performance in 1920 $\times$ 1072 Resolution CGH Generation on DIV2K/Flickr2K}
\label{table1}
\setlength{\tabcolsep}{0.5mm}
\footnotesize
\begin{tabular}{cccccc}
\toprule
\textbf{Methods} & \textbf{Parameters}& \textbf{FLOPs(G)} & \textbf{Time(ms)} & \textbf{PSNR(dB)} & \textbf{SSIM} \\ 
\midrule 
Holo-encoder & 743,896 & 35.64 & 16.66 & 23.79/23.72 & 0.613/0.614 \\ 
HoloNet & 2,868,466 &328.53& 56.59 & 28.19/27.47 & 0.803/0.816 \\ 
CCNN-CGH & 42,260 &7.21& \textbf{14.52} & 31.46/31.72 & 0.844/0.863 \\ 
Ours & \textbf{5,318} &\textbf{5.63}& 16.45 & \textbf{33.50/33.53} & \textbf{0.921/0.928} \\ 
\bottomrule
\end{tabular}
\end{table}

\begin{table}[!t]

\centering
\caption{Performance in 1920 $\times$ 1072 Resolution CGH Generation in different pixel pitches}
\label{table2}
\renewcommand{\arraystretch}{1.5}
\setlength{\tabcolsep}{3mm}
\footnotesize
\begin{tabular}{cccc}
\toprule
\textbf{Methods} & \textbf{Pixel Pitch($\mu$m)} & \textbf{PSNR(dB)} & \textbf{SSIM} \\ 
\midrule 
Holo-encoder & 3.74/6.4 & 22.91/23.37 & 0.636/0.625 \\ 
HoloNet & 3.74/6.4 & 28.33/27.41 & 0.827/0.789 \\ 
CCNN-CGH & 3.74/6.4 & 32.15/31.67 & 0.863/0.848 \\ 
Ours & 3.74/6.4 & \textbf{33.15/33.45} & \textbf{0.899/0.910} \\ 
\bottomrule
\end{tabular}
\end{table}

\begin{table}[!t]

\centering
\caption{Simulation performance of ablation study}
\label{table3}
\renewcommand{\arraystretch}{1.5}
\setlength{\tabcolsep}{1.5mm}
\footnotesize
\begin{tabular}{cccccc}
\toprule
\textbf{Methods} & \textbf{Parameters} &\textbf{FLOPs(G)}& \textbf{Time(ms)} & \textbf{PSNR(dB)} & \textbf{SSIM} \\ 
\midrule 
RC & 1,439,479 & 169.74 & 43.67 & \textbf{34.52} & 0.915 \\ 
CC & 39,296 & 9.84 & 23.49 & 32.77 & 0.889 \\ 
ND & 5,588 & 6.26 & 16.90 & 31.74 & 0.871 \\ 
NA & \textbf{5,318} & \textbf{5.63} & \textbf{15.66} & 20.66 & 0.512 \\ 
Ours & \textbf{5,318} & \textbf{5.63} & 16.45 & 33.50 & \textbf{0.921} \\ 
\bottomrule
\end{tabular}
\end{table}

\begin{table}[!t]

\centering
\caption{Performance in different channel numbers and kernel sizes}
\label{table4}
\renewcommand{\arraystretch}{1.5}
\setlength{\tabcolsep}{6mm}
\footnotesize
\begin{tabular}{cccc}
\toprule
\textbf{Channel} & \textbf{Kernel} & \textbf{PSNR(dB)} & \textbf{SSIM} \\ 
\midrule
10 & 3 & \textbf{33.71} & \textbf{0.925} \\
6 & 3 & 33.20 & 0.912 \\
8 & 3 & 33.50 & 0.921 \\ 
8 & 5 & 33.05 & 0.903 \\ 
8 & 7 & 33.18 & 0.906 \\ 
\bottomrule
\end{tabular}
\end{table}

\section{Optical experiment}

\begin{figure}[h]
\centering
\includegraphics[scale=0.8]{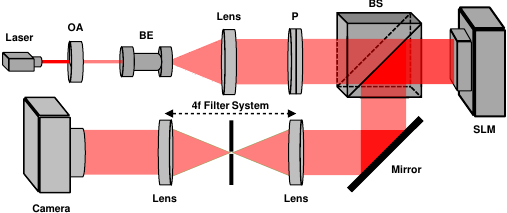}
\caption{The setup of holographic display. OA: Optical Attenuator, BE: Beam Expander, P: Polarizer, BS: Beam Splitter.}
\label{setup}
\end{figure}

Our holographic display setup is shown in Fig. \ref{setup}.  Coherent light is generated by a laser, passed a optical attenuator (OA) and beam expander (BE), then collimated using a lens. A beam splitter (BS) is employed to modify the optical path. The POH is uploaded to the SLM, which reflects and modulates the incoming light. To filter out higher diffraction orders from the holographic reconstruction, a 4f system is used, consisting of two lenses and a filter. The resolution of the phase-type SLM (FSLM-2K70-P03) used is 1920 $\times$ 1080, and the pixel pitch of it is 8 $\mu$m. Other parameters is the same as those of the numerical simulation.

\begin{figure*}[!t]
\centering
\includegraphics[scale=0.55]{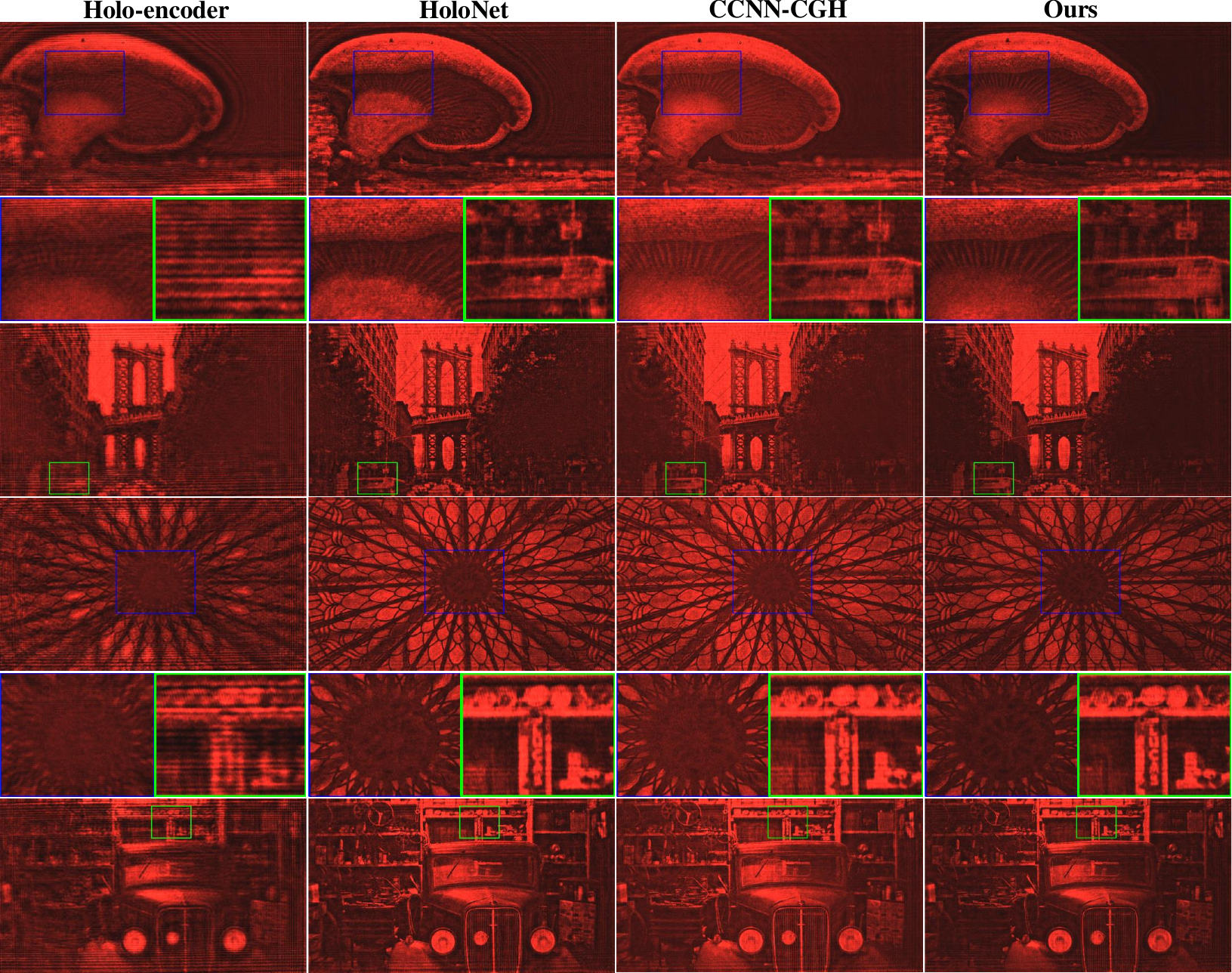}
\caption{Captured optical results of all evaluated methods.}
\label{optic}
\end{figure*}

The results of the optical experiment are presented in Fig. \ref{optic}. It is clear that the Holo-encoder performs significantly worse than the other models, as it fails to reconstruct detailed information effectively. While HoloNet offers more details compared to the Holo-encoder, it introduces blurring, leading to less clear images. Among the three comparison baseline models, CCNN-CGH shows the highest quality, but suffers from stray light and noise issues. In contrast, our model delivers more consistent reconstruction quality than CCNN-CGH, especially in terms of preserving details.

\section{Ablation study}
We conducted an ablation study using different models to evaluate performance. Our model demonstrated perfect overall performance across almost assessed metrics, thus validating its effectiveness. As shown in Table \ref{table3}, the evaluated models include those where the second model of HoloNet (RC) or CCNN-CGH (CC) is replaced with our proposed model. Additionally, to maintain a comparable number of parameters, we substituted the deformable convolution with four layers of complex-valued convolution (ND). We also move the first forward ASM (NA) to valid the effectiveness of complex amplitude as input rather than only amplitude.

After integrating our model into existing networks, all models exhibited improved performance, indicating that the use of deformable convolution enhances the quality of the reconstructed images. Specifically, the PSNR of RC reached 34.52 dB, which is 3.06 dB higher than that of CCNN-CGH. However, it has a significantly larger number of parameters and the longest inference time. Furthermore, NA shows the lowest reconstructive quality, which validates the effectiveness of using complex amplitude as input. Overall, our model strikes an optimal balance between quality and computational efficiency.

Table \ref{table4} highlights how performance varies with different initial channel numbers and kernel sizes. The model achieves its highest PSNR of 33.71 dB and SSIM of 0.925 when configured with 10 channels and a kernel size of 3. However, this setup also leads to an increase in the number of parameters and longer inference times. Furthermore, when varying the kernel sizes for deformable convolution while keeping the number of channels fixed at 8, the best results are obtained with a kernel size of 3, whereas the poorest performance is observed with a kernel size of 5.

\section{Conclusion}
In this paper, we integrate deformable convolution into complex-valued neural networks for the generation of CGH. By allowing dynamic adjustments to the convolution kernel, our approach significantly enhances the flexibility of the ERF while maintaining a shallow architecture and improving the reconstruction quality of holograms. The results demonstrate that our method not only surpasses existing open-source models such as CCNN-CGH, HoloNet, and Holo-encoder in terms of peak signal-to-noise ratio but also achieves this with a substantially reduced parameter count, making it more efficient and accessible for practical applications.

\section*{CRediT authorship contribution statement}
\textbf{Shuyang Xie:} Writing – original draft, Investigation, Software, Methodology, Validation, Formal analysis, Data curation, Conceptualization. \textbf{Jie Zhou:} Writing – review \& editing, Methodology, Validation, Formal analysis, Visualization. \textbf{Bo Xu:} Writing – review \& editing, Methodology. \textbf{Jun Wang:} Writing – review \& editing, Resource, Supervision. \textbf{Renjing Xu:} Writing – review \& editing, Supervision, Resource, Funding acquisition.

\section*{Declaration of Competing Interest}
The authors declare that they have no known competing financial interests or personal relationships that could have appeared to influence the work reported in this paper.

\section*{Acknowledgments}
This work is supported by the Guangzhou Municipal Science and Technology Project (No. 2023A03J0013).

\section*{Data Availability}
Data will be made available on request.

\bibliographystyle{elsarticle-num-names} 
\bibliography{cas-refs}





\end{document}